\documentclass[twocolumn,showpacs,prb,amsfonts,amsmath,amssymb,floatfix]{revtex4}
\usepackage{color}
\usepackage{hhline}
\usepackage{mathrsfs}
\usepackage{dcolumn}
\usepackage{multirow}
\usepackage{bm}
\usepackage{afterpage}
\usepackage{amsmath}
\usepackage{booktabs}
\usepackage{graphics}
\usepackage{graphicx}

\begin{document}

\title{Density Functional Theory for Superconductors with Particle-hole Asymmetric Electronic Structure
}

\author{Ryosuke Akashi$^{1}$}
\author{Ryotaro Arita$^{1,2}$}
\affiliation{$^1$Department of Applied Physics, The University of Tokyo, 
             Hongo, Bunkyo-ku, Tokyo 113-8656, Japan}
\affiliation{$^2$JST-PRESTO, Kawaguchi, Saitama 332-0012, Japan}

\date{\today}
\begin{abstract}
To extend the applicability of density functional theory for superconductors (SCDFT) to systems with significant particle-hole asymmetry, we construct a new exchange-correlation kernel entering the gap equation. We show that the kernel is numerically stable and does not diverge even in the low temperature limit. Solving the gap equation for model systems with the present kernel analytically and numerically, we find that the asymmetric component of electronic density of states, which has not been considered with the previous kernel, systematically decreases transition temperature ($T_{\rm c}$). We present a case where the decrease of $T_{\rm c}$ amounts to several tens of percent. 
\end{abstract}

\maketitle

\section{Introduction}
Non-empirical calculation of superconducting transition temperature ($T_{\rm c}$) has been a great challenge in theoretical physics. Conventionally, $T_{\rm c}$s of phonon-mediated superconductors are estimated by the McMillan-Allen-Dynes formula,\cite{McMillan,AllenDynes}
\[
T_{\rm c}=\frac{\omega_{\rm ln}}{1.2}\exp \left( -\frac{1.04(1+\lambda)}{\lambda-\mu^\ast(1+0.62\lambda)}\right),
\]
where $\omega_{\rm ln}$, $\lambda$ and $\mu^\ast$ (Ref.~\onlinecite{Morel-Anderson}) are the logarithmically averaged phonon frequency, electron-phonon coupling, and Coulomb pseudo-potential, respectively. This formula clearly indicates the extreme sensitivity of $T_{\rm c}$ to the phonon and electron properties; to achieve a reliable calculation of $T_{\rm c}$, we need to evaluate these factors with high accuracy. For the phonon properties, developments in methods based on density functional theory\cite{Baroni-review,Kunc-Martin-frozen} enables one to obtain reliable values very efficiently. On the other hand, for $\mu^{\ast}$, while some calculation based on the static random-phase approximation has been made,\cite{Lee-Chang-Cohen} quantitative description of the retardation effect\cite{Morel-Anderson} still remains a serious obstacle for the calculation of $T_{\rm c}$.

Recently, the situation has changed by the progress in the density-functional theory for superconductors (SCDFT).\cite{Oliveira, Kreibich, GrossI} There, a non-empirical scheme treating both the electron-phonon and electron-electron interactions was formulated referring the Migdal-Eliashberg (ME) theory of strong-coupling superconductivity.\cite{Migdal, Eliashberg, Scalapino, Schrieffer} Subsequent applications have shown that this scheme can reproduce experimental $T_{\rm c}$s of various conventional phonon-mediated superconductors with error less than a few K.\cite{GrossII, Floris-MgB2, Sanna-CaC6, Bersier-CaBeSi} 

An important advantage of the current SCDFT-based scheme is that the particle-hole symmetric component of electronic structures around the Fermi level is accurately treated on the basis of the density functional calculations. In the context of the ME theory, its effect on $T_{\rm c}$ has been intensively studied for A15 compounds.\cite{Testardi,Ho-Cohen-Pickett,Pickett-general, Lie-Carbotte, Mitrovic-Carbotte, Schachinger-Carbotte,Yokoya-Nakamura,Yokoya} However, the particle-hole antisymmetric component ignored in the current SCDFT has also been shown to have a significant impact on $T_{\rm c}$.\cite{Yokoya} Moreover, there has also been a great experimental progress of electric-field-induced carrier-doping technique, and it has stimulated current interest to superconductivity in doped insulators having significant particle-hole asymmetry.\cite{Caviglia-LAOSTO,Ueno-Kawasaki-SrTiO3, Iwasa-ZrNCl-EDLT,Ueno-Kawasaki-KTaO3,Taniguchi-MoS2,Iwasa-MoS2} With this background, it is important to extend the applicability of SCDFT to systems with particle-hole asymmetric  electronic structure.

In Ref.~\onlinecite{GrossI}, the assumption of particle-hole symmetry with respect to the Fermi energy ($E_{\rm F}$) was originally introduced to avoid numerical divergence in the exchange-correlation kernel. In this paper, we show that we can construct an exchange correlation functional for systems with asymmetric electronic structure which does not have any divergences. In Sec.~\ref{sec:theory}, we show how the kernel is constructed, and discuss how it is  numerically stable. In Sec.~\ref{sec:calc}, we perform calculations of $T_{\rm c}$ for the cases where the present improvement becomes indeed important. Section \ref{sec:summary} is devoted to summary and conclusion.

\section{gap-equation kernel for particle-hole asymmetry}\label{sec:theory}
In this section, we describe the improvement of the gap equation based on the current SCDFT to include the particle-hole asymmetry effect. In the current SCDFT,\cite{GrossI} we solve the following gap equation:
\begin{eqnarray}
\Delta_{n{\bf k}}\!=\!-\mathcal{Z}_{n\!{\bf k}}\!\Delta_{n\!{\bf k}}
\!-\!\frac{1}{2}\!\sum_{n'\!{\bf k'}}\!\mathcal{K}_{n\!{\bf k}\!n'{\bf k}'}
\!\frac{\mathrm{tanh}[(\!\beta/2\!)\!E_{n'{\bf k'}}\!]}{E_{n'{\bf k'}}}\!\Delta_{n'\!{\bf k'}}.
\label{eq:gap-eq}
\end{eqnarray}
Here, $n$ and ${\bf k}$ denote the band index and crystal momentum, respectively, $\Delta$ is the gap function, and $\beta$ is the inverse temperature. The energy $E_{n {\bf k}}$ is defined as $E_{n {\bf k}}$=$\sqrt{\xi_{n {\bf k}}^{2}+\Delta_{n {\bf k}}^{2}}$ and $\xi_{n {\bf k}}=\epsilon_{n {\bf k}}-\mu$ is the one-electron energy measured from the chemical potential $\mu$, where $\epsilon_{n {\bf k}}$ is obtained by solving the normal Kohn-Sham equation in density functional theory (DFT).

Functions $\mathcal{Z}$ and $\mathcal{K}$ are the exchange-correlation kernels, which represent the effects of mass renormalization and effective pairing interactions, respectively. In the SCDFT, $\mathcal{Z}$ is treated by only the phonon contribution as 
\begin{eqnarray}
\mathcal{Z}_{n\!{\bf k}}=\mathcal{Z}^{\rm ph}_{n\!{\bf k}}  
\end{eqnarray}
with $\mathcal{Z}^{\rm ph}_{n{\bf k}}$ representing the renormalization of electronic states by the phonon exchange, and $\mathcal{K}$ consists of both the electron-phonon $\mathcal{K}^{\rm ph}_{n\!{\bf k}\!n'{\bf k}'}$ and electron-electron $\mathcal{K}^{\rm el}_{n\!{\bf k}\!n'{\bf k}'}$ exchange contributions 
\begin{eqnarray}
\mathcal{K}_{n\!{\bf k}\!n'{\bf k}'}=
\mathcal{K}^{\rm ph}_{n\!{\bf k}\!n'{\bf k}'}+
\mathcal{K}^{\rm el}_{n\!{\bf k}\!n'{\bf k}'}.
\end{eqnarray} 

Formally, $\mathcal{Z}^{\rm ph}_{n{\bf k}}$ is derived from the functional derivative of the free energy given by the Kohn-Sham perturbation theory with respect to the anomalous density~\cite{GrossI} as
\begin{eqnarray}
\mathcal{Z}^{\rm ph,PT}_{n{\bf k}}
\equiv
\mathcal{Z}^{\rm ph}_{1,n{\bf k}}
+
\mathcal{Z}^{\rm ph}_{2,n{\bf k}}
\label{eq:Z-original}
\end{eqnarray}
with 
\begin{eqnarray}
\mathcal{Z}^{\rm ph}_{1,n{\bf k}}
&=&
\frac{1}{{\rm tanh}\bigl[(\beta/2)\xi_{n{\bf k}}\bigr]}
\sum_{n'{\bf k}'}
\sum_{\nu}
|g^{n{\bf k},n'{\bf k}'}_{\nu {\bf k}-{\bf k}'}|^{2}
\nonumber \\
&&
\times
\biggl\{\!
\frac{1}{\xi_{n{\bf k}}}
[
I(\xi_{n{\bf k}},\xi_{n'{\bf k}'},\omega_{\nu {\bf k}-{\bf k}'})\!\!
-\!\!
I(\xi_{n{\bf k}},-\xi_{n'{\bf k}'},\omega_{\nu {\bf k}-{\bf k}'})
]
\nonumber
\\
&&\hspace{40pt}
-2
I'(\xi_{n{\bf k}},\xi_{n'{\bf k}'},\omega_{\nu {\bf k}-{\bf k}'})
\biggr\}
\label{eq:Z1-original}
\end{eqnarray} 
and 
\begin{eqnarray}
\mathcal{Z}^{\rm ph}_{2,n{\bf k}}
&=&
-
\frac
{2}
{\sum_{n'{\bf k}'}(\beta/2)/{\rm cosh}^{2}
 \bigl[(\beta/2)\xi_{n'{\bf k}'} \bigr]}
\nonumber \\
&&
\times
\biggl[
\frac{1}{\xi_{n{\bf k}}}
-
\frac
{\beta/2}
{{\rm sinh\bigl[(\beta/2)\xi_{n{\bf k}}\bigr]}
{\rm cosh\bigl[(\beta/2)\xi_{n{\bf k}}\bigr]}}
\biggr]
\nonumber \label{eq:Z1-original}
\\
&&
\times \!\!
\sum_{\substack{m_{1}{\bf l}_{1}\\m_{2}{\bf l}_{2}}}\!\!
\sum_{\nu}
|g^{m_{1}{\bf l}_{1},m_{2}{\bf l}_{2}}_{\nu {\bf l}_{1}-{\bf l}_{2}}|^{2}
I'\!(\xi_{m_{1}{\bf l}_{1}},\xi_{m_{2}{\bf l}_{2}},\omega_{\nu {\bf l}_{1}-{\bf l}_{2}})
.\label{eq:Z2-original}
\end{eqnarray}
Here, $g^{n{\bf k},n'{\bf k}'}_{\nu {\bf k}-{\bf k}'}$ denotes the electron-phonon coupling, and $I$ and $I'$ are defined as
\begin{eqnarray}
I(\xi,\xi',\omega)
&=&
f_{\beta}(\xi)f_{\beta}(\xi')n_{\beta}(\omega)
\nonumber \\
&&
\biggl[
\frac{e^{\beta \xi}-e^{\beta(\xi'+\omega)}}{\xi-\xi'-\omega}
-
\frac{e^{\beta \xi'}-e^{\beta(\xi+\omega)}}{\xi-\xi'+\omega}
\biggr]
,\label{eq:I-func}
\\
I'(\xi,\xi',\omega)
&=&
\frac{\partial}{\partial \xi}
I(\xi,\xi',\omega), \label{eq:Idiff-func}
\end{eqnarray}
with $f_{\beta}$ and $n_{\beta}$ denoting the Fermi and Bose distribution functions, respectively. We note that $\mathcal{Z}^{\rm ph}_{1}$ originates from explicit dependence of the free energy on the anomalous density, whereas $\mathcal{Z}^{\rm ph}_{2}$ comes from the implicit dependence via the chemical potential.

L\"uders \textit{et al.}\cite{GrossI} reported that this form suffers from numerical divergence and that the divergent contribution is antisymmetric with respect to $E_{\rm F}$. Accordingly, in their paper they symmetrized $\mathcal{Z}^{\rm ph}_{1}$ in $\xi_{n{\bf k}}$, and dropped $\mathcal{Z}^{\rm ph}_{2}$ because of its antisymmetric dependence on $\xi_{n{\bf k}}$. The resulting form is
\begin{eqnarray}
\!\!\!\!\mathcal{Z}^{\rm ph,sym}_{n{\bf k}}
&=&
-\frac{1}{{\rm tanh}\bigl[(\beta/2) \xi_{n{\bf k}}\bigr]}
\sum_{n'{\bf k}'}
\sum_{\nu}
|g^{n{\bf k},n'{\bf k}'}_{\nu {\bf k}-{\bf k}'}|^{2}
\nonumber \\
&& \!
\times \!
\bigl[\!
I'\!(\xi_{n{\bf k}},\xi_{n'{\bf k}'}\!,\omega_{\nu {\bf k}\!-\!{\bf k}'})\!
\!+\!\!
I'\!(\xi_{n{\bf k}},\!-\xi_{n'{\bf k}'}\!,\omega_{\nu {\bf k}\!-\!{\bf k}'}\!)
\!\bigr]
.
\label{eq:Z-ph-sym}
\end{eqnarray}
Further, they modified an unphysically large component of $\mathcal{Z}^{\rm ph,sym}_{n{\bf k}}$ around $E_{\rm F}$, and obtained the following form 
\begin{eqnarray}
\!\!\!
\mathcal{Z}^{\rm ph}_{n{\bf k}}
&=&
-\frac{1}{{\rm tanh}\bigl[(\beta/2) \xi_{n{\bf k}}\bigr]}
\sum_{n'{\bf k}'}
\sum_{\nu}
|g^{n{\bf k},n'{\bf k}'}_{\nu {\bf k}-{\bf k}'}|^{2}
\nonumber \\
&& \!
\times \! \!
\bigl[
J\!(\xi_{n{\bf k}},\xi_{n'{\bf k}'}\!,\omega_{\nu {\bf k}\!-\!{\bf k}'})\!
+\!
J\!(\xi_{n{\bf k}},\!-\xi_{n'{\bf k}'}\!,\omega_{\nu {\bf k}\!-\!{\bf k}'})
\bigr]
,
\label{eq:Z-ph-prev}
\end{eqnarray}
where $J$ is defined by Eq.~(80) in Ref.~\onlinecite{GrossI}. So far, only this form has been used in practice.

In Eqs.~(\ref{eq:Z-ph-sym}) and (\ref{eq:Z-ph-prev}), the antisymmetric parts of the electronic density of states (DOS) and $|g^{n{\bf k},n'{\bf k}'}_{\nu {\bf k}-{\bf k}'}|^{2}$ with respect to $E_{\rm F}$ within the phonon energy scale are ignored. Within the standard ME theory, the effect of their antisymmetric parts is assumed to be minor\cite{Pickett-general, Mitrovic} and often ignored for simplicity.~\cite{Schrieffer} On the other hand, in systems with rapidly varying DOS, this treatment is also known to lead to a considerable error in the estimate of $T_{\rm c}$ and the isotope-effect coefficient.\cite{Yokoya} Below, we will show a numerically stable form of $\mathcal{Z}$ without the symmetrization, with which the asymmetry effect is properly treated.

\subsection{Cancellation of divergent terms}\label{subsec:cancel}
First, we analytically examine the divergence of $\mathcal{Z}^{\rm ph,PT}$. For simplicity, we deal with the $n{\bf k}$-averaged form of $\mathcal{Z}^{\rm ph,PT}$, defined by
\begin{eqnarray}
\mathcal{Z}^{\rm ph,PT}(\xi)
&=&
\mathcal{Z}^{\rm ph}_{1}(\xi)
+
\mathcal{Z}^{\rm ph}_{2}(\xi)
,\label{eq:Z-ave-original}
\\
\mathcal{Z}^{\rm ph}_{i}(\xi)
&\equiv&
\frac{1}{N(\xi)}
\sum_{n{\bf k}} \delta(\xi-\xi_{n{\bf k}}) \mathcal{Z}^{\rm ph}_{i,n{\bf k}}
\hspace{5pt} (i=1,2)
,
\label{eq:Z-ave-Z1Z2}
\end{eqnarray}
where $N(\xi)$ is the electronic DOS. We also introduce an approximation for the electron-phonon coupling
\begin{eqnarray}
&&
\sum_{n{\bf k} n'{\bf k}' \atop \nu} \!\!\!\!
\delta(\xi \!-\! \xi_{n{\bf k}})
\delta(\xi' \!-\! \xi_{n'{\bf k}'})
\delta(\omega \!-\! \omega_{\nu{\bf k}-{\bf k}'})
|g^{n{\bf k},n'{\bf k}'}_{\nu {\bf k}-{\bf k}'}|^{2}
\nonumber \\
&&\hspace{50pt}
\simeq
\frac{N(\xi)N(\xi')}{N(0)} \alpha^{2}F(\omega)
,
\label{eq:G2-approx2}
\end{eqnarray}
where the antisymmetric component of DOS is retained. The function $\alpha^{2}F(\omega)$ denotes the Eliashberg function, with which the electron-phonon coupling coefficient $\lambda$ and the characteristic frequency $\omega_{\rm ln}$ are defined as
\begin{eqnarray}
\lambda
&=&
2\int d\omega
\frac{
\alpha^{2}F(\omega)
}{\omega}
,
\label{eq:lambda-def}
\\
\omega_{\rm ln}
&=&
{\rm exp}\left[
\frac{2}{\lambda}
\int d\omega
\frac{
\alpha^{2}F(\omega)
}{\omega}
{\rm ln}\omega
\right]
.
\label{eq:omega-def}
\end{eqnarray}

Starting from the above form for $\mathcal{Z}$, in brief, we show that the numerically divergent terms analytically cancel with each other so that we can construct a numerically stable form. Readers who are not interested in the detail of this analysis can skip the remaining part of this subsection.

Let us first focus on $\mathcal{Z}^{\rm ph}_{1}$. As we show below, the divergence in this part basically comes from the slow decay of order $O(\xi'^{-1})$ of the integrands. In order to extract this, we transform the following factor appearing in $I$ and $I'$ [Eqs.~(\ref{eq:I-func}) and (\ref{eq:Idiff-func})] as
\begin{eqnarray}
\frac{1}{\xi-\xi'\pm\omega}
=
{\rm P}
\frac{\xi}{(\xi-\xi'\pm\omega)(\xi'\mp\omega)}
-
{\rm P}
\frac{1}{\xi'\mp\omega}
,
\end{eqnarray}
The principal value integral ${\rm P}$ has been introduced to deal with the poles at $\xi'$$=$$\pm \omega$ in the right hand side; note that the expression in the left hand side does not have these poles. Substituting this expression into Eq.~(\ref{eq:Z-ave-Z1Z2}), inserting identities $1=\int d\xi' \delta(\xi'-\xi_{n'{\bf k}'})$ and $1=\int d\omega \delta(\omega-\omega_{\nu{\bf k}-{\bf k}'})$, and using Eq.~(\ref{eq:G2-approx2}), we obtain
\begin{widetext}
\begin{eqnarray}
\mathcal{Z}^{\rm ph}_{1}(\xi)
&=&
\frac{1}{{\rm tanh}[(\beta/2)\xi]}
\int d\omega 
\int d\xi'
\frac{N(\xi')}{N(0)} \alpha^{2}F(\omega)
\biggl\{
I_{0}(\xi,\xi',\omega)
+
I_{1}(\xi,\xi',\omega)
\nonumber \\
&& \hspace{70pt}
-2
[J_{0}(\xi,\xi',\omega)
 +J_{1}(\xi,\xi',\omega)
 +J_{2}(\xi,\xi',\omega)
 ]
\biggr\}
, \label{eq:Z-IJ-def}
\\
I_{0}(\xi,\xi',\omega)
&=&
-
\frac{1}{\xi}
f_{\beta}(\xi)f_{\beta}(\xi')n_{\beta}(\omega)
\bigl[
{\rm P}\frac{e^{\beta \xi} \!-\! e^{\beta(\xi'+\omega)}}{\xi'+\omega}
\!-\!
{\rm P}\frac{e^{\beta \xi'} \!-\! e^{\beta(\xi+\omega)}}{\xi'-\omega}
\!-\!
{\rm P}\frac{1 \!-\! e^{\beta(\xi+\xi'+\omega)}}{\xi'+\omega}
\!+\!
{\rm P}\frac{e^{\beta (\xi+\xi')} \!-\! e^{\beta(\omega)}}{\xi'-\omega}
\bigr]
,\label{eq:I0-def}
\\
I_{1}(\xi,\xi',\omega)
&=&
f_{\beta}(\xi)\!f_{\beta}(\xi'\!)\!n_{\beta}(\omega\!)\!
\biggl[\!
{\rm P}\frac{e^{\beta \xi}-e^{\beta(\xi'+\omega)}}{(\xi \!\!-\!\! \xi' \!\!\!-\!\! \omega)(\xi' \!\!\!+\!\! \omega)}
\!-\!
{\rm P}\frac{e^{\beta \xi'}-e^{\beta(\xi+\omega)}}{(\xi \!\!-\!\! \xi' \!\!\!+\!\! \omega)(\xi' \!\!\!-\!\! \omega)}
\!-\!
{\rm P}\frac{1-e^{\beta(\xi+\xi'+\omega)}}{(\xi \!\!+\!\! \xi' \!\!\!+\!\! \omega)(\xi' \!\!\!+\!\! \omega)}
\!+\!
{\rm P}\frac{e^{\beta (\xi+\xi')}-e^{\beta \omega}}{(\xi \!\!+\!\! \xi' \!\!\!-\!\! \omega)(\xi' \!\!\!-\!\! \omega)}
\!
\biggr],
\label{eq:I1-def}
\\
J_{0}(\xi,\xi',\omega)
&=&
f'_{\beta}(\xi)f_{\beta}(\xi')n_{\beta}(\omega)
\bigl[
{\rm P}\frac{1+e^{\beta(\xi'+\omega)}}{\xi'+\omega}
+
{\rm P}\frac{e^{\beta \xi'}+e^{\beta \omega}}{\xi'-\omega}
\bigr]
,\label{eq:J0-def}
\\
J_{1}(\xi,\xi',\omega)
&=&
-f'_{\beta}(\xi)f_{\beta}(\xi')n_{\beta}(\omega)\xi
\biggl[
{\rm P}\frac{1+e^{\beta(\xi'+\omega)}}{(\xi-\xi'-\omega)(\xi'+\omega)}
+
{\rm P}\frac{e^{\beta\xi'}+e^{\beta \omega}}{(\xi-\xi'+\omega)(\xi'-\omega)}
\biggr]
,
\label{eq:J1-def}
\\
J_{2}(\xi,\xi',\omega)
&=&
-f_{\beta}(\xi)f_{\beta}(\xi')n_{\beta}(\omega)
\biggl[
\frac{e^{\beta \xi}-e^{\beta(\xi'+\omega)}}{(\xi-\xi'-\omega)^{2}}
-
\frac{e^{\beta \xi'}-e^{\beta(\xi+\omega)}}{(\xi-\xi'+\omega)^{2}}
\biggr]
\label{eq:J2-def}
.
\end{eqnarray}
\end{widetext}
Here $I_{i}$ ($i$$=$$0,1$) and $J_{i}$ ($i$$=$$0,1,2$) denote the terms originating from $I$ and $I'$, respectively. From this expression, the divergent contribution $\mathcal{Z}^{\rm ph,div}_{1}$ is extracted as
\begin{eqnarray}
\mathcal{Z}^{\rm ph,div}_{1}(\xi)
\!&=&\!
\frac{1}{{\rm tanh}[(\beta/2)\xi]}
\int d\omega 
\int d\xi'
\frac{N(\xi')}{N(0)} \alpha^{2}F(\omega)
\nonumber \\
&&\hspace{20pt}
\times
\biggl\{
I_{0}(\xi,\xi'\!,\omega)
\!-\!
2J_{0}(\xi,\xi'\!,\omega)
\biggr\}
.\label{eq:Z1-div}
\end{eqnarray}
We do not explicitly write the variables $(\xi,\xi',\omega)$ unless necessary in the following.

\begin{figure}[b!]
\begin{center}
\includegraphics[width=6cm,clip]{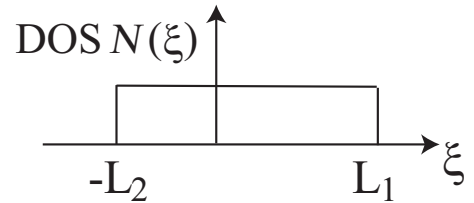}
\end{center}
\caption{Constant DOS with asymmetric cutoff energies, $L_{1}$ and $-L_{2}$.}
\label{fig:DOS-const}
\end{figure}

Let us examine $\mathcal{Z}^{\rm ph,div}_{1}$. For simplicity, we start from the case of constant electronic DOS $N(\xi)$$=$$N(0)$ with asymmetric energy cutoffs $[-L_{2},L_{1}]$ as depicted in Fig.~\ref{fig:DOS-const}, which is the simplest case of particle-hole asymmetry. We then get to

\begin{eqnarray}
\mathcal{Z}^{\rm ph,div}_{1}(\xi)
\!=\!
\frac{1}{{\rm tanh}[(\beta/2)\xi]}
\!\int \!\!d\omega \alpha^{2}F(\omega) 
\int^{L_{1}}_{-L_{2}} \!\!d\xi'
\nonumber \\
\hspace{20pt}\times
\biggl\{
I_{0}
\!-\!
2J_{0}
\biggr\}
.
\end{eqnarray}

Next we carry out the $\xi'$ integral. Since the integrand is of order $O(\xi'^{-1})$, integration results in cutoff-dependent terms as
\begin{eqnarray}
\int^{L_{1}}_{-L_{2}} d\xi' I_{0}
&\simeq&
\frac{1}{\xi}
\bigl[
f_{\beta}(\xi)
-
f_{\beta}(-\xi)
\bigr]
  {\rm ln}\left|\frac{L_{1}+\omega}{L_{2}+\omega}\right|
,
\\
\int^{L_{1}}_{-L_{2}} d\xi' J_{0}
&\simeq&
f'_{\beta}(\xi)
 {\rm ln}\left|\frac{L_{1}+\omega}{L_{2}+\omega}\right|
.
\end{eqnarray}
Here we have neglected terms of order $O(e^{-\beta \omega})$, $O(e^{-\beta L_{1}})$ and $O(e^{-\beta L_{2}})$, and this approximation is implicitly used in the following equations with ``$\simeq$" in this section. We consequently get to
\begin{eqnarray}
\mathcal{Z}^{\rm ph,div}_{1}(\xi)
&\simeq&
-
\left\{
\frac{1}{\xi}
-
\frac{\beta/2}{{\rm sinh[(\beta/2)\xi]}{\rm cosh[(\beta/2)\xi]}}
\right\}
\nonumber \\
&& \hspace{35pt}
\times
\int d\omega
\alpha^{2}F(\omega)
{\rm ln}\left|\frac{L_{1}+\omega}{L_{2}+\omega}\right|
.
\label{eq:I0J0-final}
\end{eqnarray}
The $\xi$-dependent factor here is reminiscent of $\mathcal{Z}^{\rm ph}_{2}$ [Eq.~(\ref{eq:Z2-original})]. In fact, we prove that this term exactly cancels $\mathcal{Z}^{\rm ph}_{2}(\xi)$ in the following.

Next, we consider $\mathcal{Z}^{\rm ph}_{2}(\xi)$ using the approximations of the same level. The $n{\bf k}$-averaged form is explicitly written as
\begin{eqnarray}
\mathcal{Z}^{\rm ph}_{2}(\xi)
&=&
-
\biggl[
\frac{1}{\xi}
-
\frac
{\beta/2}
{{\rm sinh\bigl[(\beta/2)\xi\bigr]}
 {\rm cosh\bigl[(\beta/2)\xi\bigr]}}
\biggr]
\nonumber
\\
&& \!\!\!\!\!\!\!\!
\times \!\!
\int \! d\omega \alpha^{2}F(\omega) \! \!
\int^{L_{1}}_{-L_{2}}\!\!\!d\xi' \! 
\int^{L_{1}}_{-L_{2}}\!\!\!d\xi  \!
I'(\xi,\xi',\omega)
,
\end{eqnarray}
where we have used
\begin{eqnarray}
\sum_{n'{\bf k}'}(\beta/2)/{\rm cosh}^{2}
 \bigl[(\beta/2)\xi_{n'{\bf k}'} \bigr]
\simeq
2N(0)
.
\end{eqnarray}
By carrying out the integration $\int d\xi \int d\xi'$ and ignoring terms of order $O[(T/L_{1})^{2},(T/L_{2})^{2}]$, we get
\begin{eqnarray}
\int^{L_{1}}_{-L_{2}} d \xi'
\int^{L_{1}}_{-L_{2}} d \xi
I'(\xi,\xi',\omega)
\simeq
{\rm ln}\left|\frac{L_{2}+\omega}{L_{1}+\omega} \right|
,
\end{eqnarray}
and immediately obtain
 \begin{eqnarray}
\mathcal{Z}^{\rm ph}_{2}(\xi)
&\simeq&
\biggl[
\frac{1}{\xi}
-
\frac
{\beta/2}
{{\rm sinh\bigl[(\beta/2)\xi\bigr]}
 {\rm cosh\bigl[(\beta/2)\xi\bigr]}}
\biggr]
\nonumber
\\
&&
\times
\int d\omega \alpha^{2}F(\omega)
{\rm ln}\left|\frac{L_{1}+\omega}{L_{2}+\omega} \right|
.
\label{eq:Z2-final}
\end{eqnarray}
This expression exactly cancels out $\mathcal{Z}^{\rm ph,div}_{1}(\xi)$ [Eq.~(\ref{eq:I0J0-final})].

One can observe the same cancellation for the nonconstant DOS case. The cutoff-dependent parts  $\mathcal{Z}^{\rm ph,div}_{1}$ and $\mathcal{Z}^{\rm ph}_{2}$ then read
\begin{eqnarray}
\mathcal{Z}^{\rm ph,div}_{1}(\xi)
&\simeq&
\frac{1}{{\rm tanh}\frac{\beta}{2}\xi}\!
\int \!\! d\omega a^{2}F(\omega) \!\!
\int \!\! d\xi' \frac{N(\xi')}{N(0)}
\bigl[\!
I_{0}
\!-\!2J_{0}
\!\bigr]
,\label{eq:Z1-nonconst}
\\
\mathcal{Z}^{\rm ph}_{2}(\xi)
&\simeq&
-
\biggl\{\!
\frac{1}{\xi}
\!-\!
\frac{\beta/2}{{\rm cosh}[(\beta/2)\xi]{\rm sinh}[(\beta/2)\xi]}
\biggr\}\!\!
\int \!\!d\omega
\alpha^{2}\!F(\omega)
\nonumber \\
&&
\times
\int^{L_{1}}_{-L_{2}}\!\!\!d\xi
\frac{N(\xi)}{N(0)}
\int^{L_{1}}_{-L_{2}}\!\!\!d\xi'
\frac{N(\xi')}{N(0)}
I'(\xi,\xi',\omega)
.
\label{eq:Z2-nonconst}
\end{eqnarray}
\begin{figure}[t!]
\begin{center}
\includegraphics[width=8cm,clip]{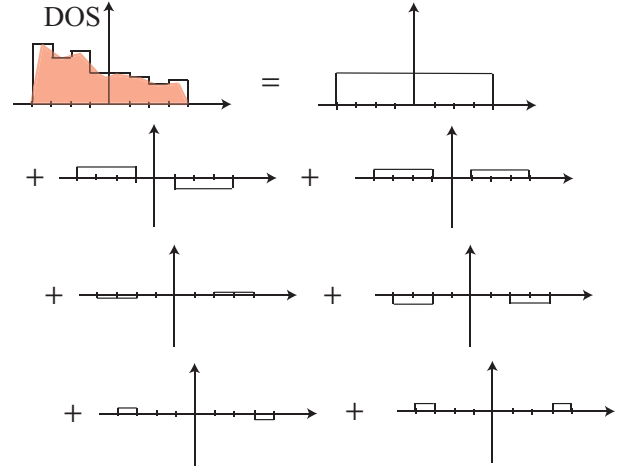}
\end{center}
\caption{Decomposition procedure of DOS. Typical DOS is represented as red-shaded area.}
\label{fig:DOS-general}
\end{figure}

Here, we approximate the DOS as a step-like function with a certain set of energy tics $\{\pm\epsilon_{i}\}$ $(i=1, 2, . . ., N_{\epsilon}; \epsilon_{i} < {\rm min}[L_{1},L_{2}])$. The approximate function is then decomposed into functions $\theta^{+}_{i}(\xi) \equiv \theta(\xi-\epsilon_{i}) + \theta(-\xi-\epsilon_{i})$ and $\theta^{-}_{i}(\xi) \equiv \theta(\xi-\epsilon_{i}) - \theta(-\xi-\epsilon_{i})$ as
\begin{eqnarray}
N(\xi)
=
N(0)
\biggl\{
1+
\sum_{i=1}^{N_{\epsilon}}
[
N^{-}_{i}\theta^{-}_{i}(\xi)
+
N^{+}_{i}\theta^{+}_{i}(\xi)
]
\biggr\}
.
\label{eq:DOS-step}
\end{eqnarray}
This procedure is schematically depicted in Fig.~\ref{fig:DOS-general}. For $\mathcal{Z}^{\rm ph,div}_{1}$, straightforward calculations yield
\begin{eqnarray}
&&
\mathcal{Z}^{\rm ph,div}_{1}\!(\xi)
\!\simeq\!
-
\biggl\{\!
\frac{1}{\xi}
\!-\!
\frac{\beta/2}{{\rm cosh}[(\beta/2)\xi]{\rm sinh}[(\beta/2)\xi]}
\biggr\}\!\!\!
\int \!\! d\omega \alpha^{2}\!F(\omega)
\nonumber \\
&& \hspace{40pt}
\times
\biggl\{
{\rm ln}
\left|
\frac{L_{1} \!+\! \omega}{L_{2}\!+\!\omega}
\right|
\!+\!
\sum_{i}
\biggl[
N^{-}_{i}
{\rm ln}
\left|
\frac{(L_{1} \!+\! \omega)(L_{2} \!+\! \omega)}{(\epsilon_{i} \!+\! \omega)^{2}}
\right|
\nonumber \\
&& \hspace{100pt}
+
N^{+}_{i}
{\rm ln}
\left|
\frac{L_{1} \!+\! \omega}{L_{2} \!+\! \omega}
\right|
\biggr]
\biggr\}
.
\label{eq:Z1-generalDOS}
\end{eqnarray}
For $\mathcal{Z}^{\rm ph}_{2}$, terms proportional to $N^{\pm}_{i}N^{\pm}_{j}$ (any double sign) seemingly become nonzero, but careful calculations show that these terms cancel out each other (See Appendix \ref{app:eval-Z2}). Finally, we see that the cutoff-dependent part in $\mathcal{Z}^{\rm ph}_{2}$ exactly cancels Eq.~(\ref{eq:Z1-generalDOS}). By taking limit for the number of tics $N_{\epsilon} \rightarrow \infty$, the present cancellation is proved even for general DOS.\cite{note-proof}

\begin{figure}[t!]
\begin{center}
\includegraphics[width=9cm,clip]{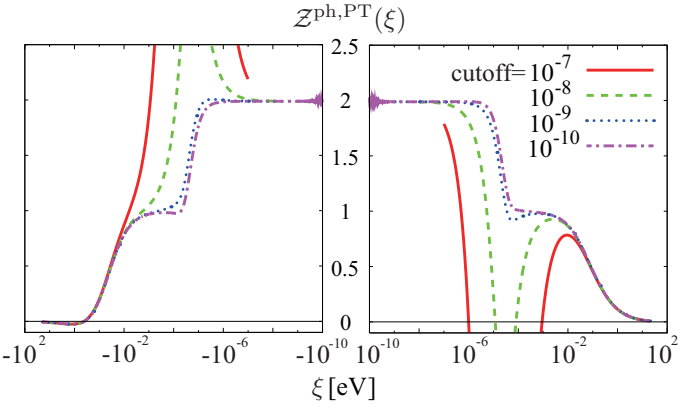}
\end{center}
\caption{Calculated values of $\mathcal{Z}^{\rm ph,PT}(\xi)$ [Eq.~(\ref{eq:Z-ave-original})] with the approximation given in Eq.~\ref{eq:G2-approx2}. Model DOS depicted in Fig.~\ref{fig:Tc-ratio} with $E_{\rm F}=0$ and model Eliashberg function given by Eq.~(\ref{eq:a2F-model}) with $\lambda=1$, $\omega_{E}=400$cm$^{-1}$, and $\sigma=\omega_{E}/10$ were used. The calculations were done with $T=0.1$K. Logarithmically uniform energy grids with mininum energy cutoff of $10^{-7}$ (red solid line), $10^{-8}$ (green dashed line), $10^{-9}$ (blue dotted line), and $10^{-10}$ (violet, dot-dashed line) were used. The DOS values for $|\xi|$$<$$10T$ were fixed to $N(0)$ so that it is consistent with the condition ($\epsilon_{i}$$\gg$$T$) assumed in Sec.~\ref{subsec:cancel}.}
\label{fig:Zconv}
\end{figure}

Now let us point out the important aspect of Eq.~(\ref{eq:Z-ave-original}): With the present approximation, the cutoff dependent divergence cancels if analytically calculated, and the apparently divergent contribution is in fact nonsingular with respect to $\xi$, since it is proportional to a nonsingular function 
$
\frac{1}{\xi}
-
\frac{\beta/2}{{\rm cosh}[(\beta/2)\xi]{\rm sinh}[(\beta/2)\xi]}
$.
In order to verify this, we have calculated $\mathcal{Z}^{\rm ph,PT}(\xi)$ [Eq.~(\ref{eq:Z-ave-original})] with a certain asymmetric model DOS [depicted later in Sec.~\ref{sec:calc}]. We have generated the energy grid so that it becomes uniform in logarithmic scale. The calculated result is shown in Fig.~\ref{fig:Zconv}. By lowering the minimum energy cutoff for the grid, convergence within order of $\lambda$ has been achieved, where the apparent divergence gradually vanishes. However, we have also found that the convergence requires formidably accurate integration within the energy scale smaller than temperature, and so it is extremely difficult to achieve in practical calculations.

\subsection{$\mathcal{Z}$ kernel for particle-hole asymmetic systems}\label{subsec:derive-Z-new}
The above analysis specifies the numerically diverging (or \textit{unstable}) but analytically irrelevant terms. In order to circumvent the numerical difficulty, we propose to simply subtract $\mathcal{Z}^{\rm ph,div}_{1}$ and $\mathcal{Z}^{\rm ph}_{2}$, obtaining
\begin{eqnarray}
\mathcal{Z}^{\rm ph,aux}(\xi)
&=&
\frac{1}{{\rm tanh}[(\beta/2)\xi]}
\int d\omega
\alpha^{2}F(\omega)
\int d\xi'
\frac{N(\xi')}{N(0)}
\nonumber \\
&&
\times
\bigl[
I_{1}
-2J_{1}-2J_{2}
\bigr]
.
\label{eq:Z-aux}
\end{eqnarray}

This auxiliary expression shows stable convergence, includes the effect of the antisymmetric part of electronic states, and agrees with Eq.~(\ref{eq:Z-ave-original}) with error of order $O[(T/L_{1})^{2}, (T/L_{2})^{2}]$ in the limit where Eq.~(\ref{eq:G2-approx2}) becomes exact. The form given in Eq.~(\ref{eq:Z-aux}) is a generalization of the $n{\bf k}$-averaged form of Eq.~(\ref{eq:Z-ph-sym}) 
\begin{eqnarray}
\mathcal{Z}^{\rm ph,sym}(\xi)
&=&
-\frac{1}{{\rm tanh}\bigl[(\beta/2) \xi\bigr]}
\int d\omega \alpha^{2}F(\omega)
\int d\xi'
\frac{N(\xi')}{N(0)}
\nonumber \\
&& \!
\times \!
\bigl[
I'\!(\xi,\xi',\omega)\!
+\!
I'\!(\xi,\!-\xi',\omega)
\bigr]
:
\label{eq:Z-ph-ave-sym}
\end{eqnarray}
Symmetrization of Eq.~(\ref{eq:Z-aux}) in $\xi$ yields Eq.~(\ref{eq:Z-ph-ave-sym}), the limiting value $\lim_{\xi \rightarrow 0}[\mathcal{Z}^{\rm ph,aux}(\xi)-\mathcal{Z}^{\rm ph,sym}(\xi)]=0$ because of the fact that the integrand for $\xi$$=$$0$ is symmetric in $\xi'$, and the two forms show almost the same temperature dependence.

Following the procedure of Ref.~\onlinecite{GrossI}, we next modify $\mathcal{Z}^{\rm ph,aux}(\xi)$ around $E_{\rm F}$ and obtain
\begin{eqnarray}
&&
\mathcal{Z}^{\rm ph,aux2}(\xi)
\nonumber \\
&& \hspace{10pt}
=
\frac{1}{{\rm tanh}[(\beta/2)\xi]}
\int d\omega
\alpha^{2}F(\omega)
\int d\xi'
\frac{N(\xi')}{N(0)}
\nonumber \\
&& \hspace{30pt}
\times
\bigl[
I_{1}
\!-\! 2J'_{1}\!-\! 2J_{2}
\bigr]
,
\label{eq:Z-l2-def}
\\
&&
J'_{1}
=
f'_{\beta}(\xi)\xi
\biggl[ \!
\bigl\{
\! f_{\beta}(\xi \!-\! \omega) \!+\! n_{\beta}(-\omega) \!
\bigr\}
{\rm P}\frac{1}{(\xi \!-\! \xi' \!-\! \omega)(\xi' \!+\! \omega)}
\nonumber \\
&& \hspace{30pt}
-
\bigl\{
f_{\beta}(\xi \!+\! \omega) \!+\! n_{\beta}(\omega)
\bigr\}
{\rm P}\frac{1}{(\xi \!-\! \xi' \!+\! \omega)(\xi' \!-\! \omega)}
\biggr]
.
\label{eq:J1p-def}
\end{eqnarray}

In practice, we also found that the principal-value integral around the singularity at $\xi'=\pm\omega$ is numerically unstable and its convergence is difficult to achieve with a practical computational cost. We therefore introduce an even smoothing function $p(\xi'\pm\omega)$ whose value is unity for $|\xi'\pm\omega| \gtrsim T$ and of order $O[(\xi'\pm\omega)^{2}]$ for $|\xi'\pm\omega|\ll T$. Substituting $p(\xi'\pm\omega)/(\xi'\pm\omega)$ for ${\rm P}[1/(\xi'\pm\omega)]$ and rearranging the terms yield\cite{comment-subtract}
\begin{eqnarray}
\mathcal{Z}^{\rm ph,new}(\xi)
&=&
\frac{1}{{\rm tanh}[(\beta/2)\xi]}
\!\!\int \!\! d\omega
\alpha^{2}\!F(\omega) \!\!
\int \!\! d\xi'
\frac{N(\xi')}{N(0)}
\bigl[
\mathcal{I}
\!-\! 2\mathcal{J}
\bigr]
, \nonumber \\
\label{eq:Z-ave-new}
\\
\mathcal{I}(\xi\!,\xi'\!\!,\omega)
&=&
\tilde{\mathcal{I}}(\xi\!,\xi'\!\!,\omega)
\!-\!
\tilde{\mathcal{I}}(\xi\!,\xi'\!\!,\!-\omega)
\nonumber \\
&& \hspace{40pt}
\!-\!
\tilde{\mathcal{I}}(\xi\!,\!-\xi'\!\!,\omega)
\!+\!
\tilde{\mathcal{I}}(\xi\!,-\xi'\!\!,\!-\omega)
,\label{eq:mathcalI-def}
\\
\tilde{\mathcal{I}}(\xi\!,\xi'\!\!,\omega)
&=&
[f_{\beta}(\xi)\!+\!n_{\beta}(\omega)]\frac{f_{\beta}(\xi')\!\!-\!\!f_{\beta}(\xi\!-\!\omega)}{\xi\!-\!\xi'\!-\!\omega}
\frac{p(\xi'\!\!+\!\omega)}{\xi'\!\!+\!\omega}
,\label{eq:tildeI-def}
\\
\mathcal{J}(\xi\!,\xi'\!\!,\omega)
&=&
\tilde{\mathcal{J}}(\xi\!,\xi'\!\!,\omega)
\!-\!
\tilde{\mathcal{J}}(\xi\!,\xi'\!\!,-\omega)
,\label{eq:mathcalJ-def}
\\
\tilde{\mathcal{J}}(\xi\!,\xi'\!\!,\omega)
&=&
-
\frac{f_{\beta}(\xi)\!+\!n_{\beta}(\omega)}{\xi\!-\!\xi'\!-\!\omega}
p(\xi'\!\!+\!\omega)
\biggl[
\frac{f_{\beta}(\xi')\!-\!f_{\beta}(\xi\!-\!\omega)}{\xi\!-\!\xi'\!-\!\omega}
\nonumber \\
&&\hspace{40pt}
-
\beta f_{\beta}(\xi\!-\!\omega)f_{\beta}(\!-\xi\!+\!\omega)
\frac{\xi}{\xi'\!+\!\omega}
\biggr]
.\label{eq:tildeJ-def}
\end{eqnarray}
We have used $p(x)=[{\rm tanh}(500\beta x)]^{4}$ in this paper, though the calculated value is not sensitive to a specfic choice of $p(x)$. 

With the above form, the asymmetry effect is properly treated in a numerically stable manner; the smoothness of the integrands at $\xi'=\pm\xi\pm\omega, \pm \omega$, and $\xi=0$ is retained. In addition, for approximately symmetric DOS, the calculated results are guaranteed to be quite close to those with the $n{\bf k}$-averaged form of Eq.~(\ref{eq:Z-ph-prev})
\begin{eqnarray}
\mathcal{Z}^{\rm ph}(\xi)
&=&
-\frac{1}{{\rm tanh}[(\beta/2) \xi]}
\int d\omega \alpha^{2}F(\omega)
\int d\xi'
\frac{N(\xi')}{N(0)}
\nonumber \\
&& \!
\times \!
\bigl[
J\!(\xi,\xi',\omega)\!
+\!
J\!(\xi,\!-\xi',\omega)
\bigr]
\label{eq:Z-ph-ave-prev}
\end{eqnarray} 
because of the following properties: (i) $\lim_{\xi \rightarrow 0}[\mathcal{Z}^{\rm ph, new}(\xi)-\mathcal{Z}^{\rm ph}(\xi)]\simeq 0$, and particularly for constant DOS, the desirable behavior\cite{GrossI} $\lim_{\xi \rightarrow 0}\mathcal{Z}^{\rm ph,new}(\xi)\simeq\lim_{\xi \rightarrow 0}\mathcal{Z}^{\rm ph}(\xi) \simeq \lambda$ is satisfied, (ii) temperature dependence of the calculated value is similar to that of $\mathcal{Z}^{\rm ph}(\xi)$, and (iii) the symmetric part of the integrand, that is,
\begin{eqnarray}
\frac{
\mathcal{I}(\xi,\xi' \!\!,\omega)
\!+\!
\mathcal{I}(\xi,\!-\xi' \!\!,\omega)
\!-\!
2\mathcal{J}(\xi,\xi' \!\!,\omega)
\!-\!
2\mathcal{J}(\xi,\!-\xi' \!\!,\omega)
}{2}
\label{eq:sym-part}
\end{eqnarray}
agrees with $J\!(\xi,\xi',\omega)\!+\!J\!(\xi,\!-\xi',\omega)$ in Eq.~(\ref{eq:Z-ph-ave-prev}) when we impose a condition $\omega$$\gtrsim$$|\xi|$$\gg$$T$. These properties are demonstrated in the following calculations. Further, we have found that the $n{\bf k}$-resolved counterpart of Eq.~(\ref{eq:Z-ave-new}) given by
\begin{eqnarray}
\!\!\!\!\!\!\!\!
\mathcal{Z}^{\rm ph, new}_{n{\bf k}}\!\!
&=&
\frac{1}{{\rm tanh}[(\beta/2)\xi]}
\sum_{n'{\bf k}' \nu}
|g^{{\bf k}-{\bf k}'\nu}_{n{\bf k},n'{\bf k}'}|^{2}
\nonumber \\
&& \hspace{-10pt}
\times
\bigl[
\mathcal{I}\!(\xi_{n{\bf k}},\xi_{n'{\bf k}'}\!,\omega_{\nu{\bf k}-{\bf k}'}\!)
\!-\!
2\mathcal{J}\!(\xi_{n{\bf k}},\xi_{n'{\bf k}'}\!,\omega_{\nu{\bf k}-{\bf k}'}\!)
\bigr]
\label{eq:Z-new}
\end{eqnarray}
is also numerically stable, which is a reasonable generalization of Eq.~(\ref{eq:Z-ph-prev}). Equations (\ref{eq:Z-ave-new}) and (\ref{eq:Z-new}) are those we propose as the new forms of $\mathcal{Z}$ for particle-hole asymmetry.

\section{Asymmetry effect on $T_{\rm c}$}\label{sec:calc}
In the rest of the present paper we get insights into the asymmetry effect included in the present improvement. In Sec.~\ref{subsec:analytic}, we analytically construct a formula of $T_{\rm c}$ to discuss in what situation the asymmetry significantly affects $T_{\rm c}$. In Sec.~\ref{subsec:modelcalc}, a typical model case is presented where our new kernel and the previous one give quite different $T_{\rm c}$, and some remarks on the application to elemental metals are given. We here note that we focus on the $n{\bf k}$-averaged form [Eq.~(\ref{eq:Z-ave-new})], and thereby the present scope includes only the asymmetry of DOS, not that of the electron-phonon matrix elements.

\subsection{Analytic calculation}\label{subsec:analytic}
\begin{figure}[t!]
\begin{center}
\includegraphics[width=8cm,clip]{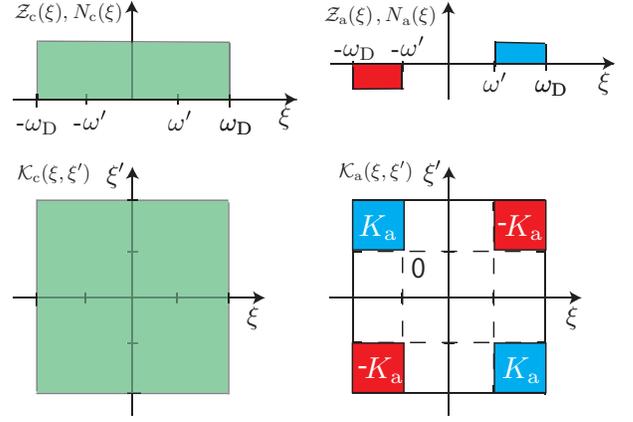}
\end{center}
\caption{(Color online) Schematic description of terms defined by Eqs.~(\ref{eq:Z-McM})--(\ref{eq:K-McM}).}
\label{fig:terms}
\end{figure}

We first analytically solve the energy-averaged gap equation\cite{GrossI}
\begin{eqnarray}
\Delta(\xi)
&=&
-\mathcal{Z}(\xi)\Delta(\xi)
\nonumber \\
&& \hspace{5pt}
-\frac{1}{2}
\int \!\! d\xi' \!
N(\xi') \mathcal{K}(\xi\!,\xi')\frac{{\rm tanh}[(\beta/2)\xi']}{\xi'}\Delta(\xi')
,
\label{eq:gap-eq-ave}
\end{eqnarray}
with the electron-phonon kernels only. We introduce a simple model for which DOS and kernels are given by the constant part and the antisymmetric part, given as $\mathcal{Z}(\xi)=\mathcal{Z}_{\rm c}(\xi) +\mathcal{Z}_{\rm a}(\xi)$, $N(\xi)=N_{\rm c}(\xi) +N_{\rm a}(\xi)$, $\mathcal{K}(\xi, \xi')=\mathcal{K}_{\rm c}(\xi,\xi') +\mathcal{K}_{\rm a}(\xi,\xi')$, with each part defined within the Debye frequency $\omega_{\rm D}$ by
\begin{eqnarray}
\mathcal{Z}_{\rm c}(\xi) 
&=&
\!Z_{\rm c}
,  \ \ 
\mathcal{Z}_{\rm a}(\xi)
\!=\!
\left\{
\begin{array}{ll}
\!-\!Z_{\rm a}, & \  ( \xi \!\leq\!-\omega' ) \\
0,     & \  (|\xi| \!<\! \omega') \\
Z_{\rm a}, & \  ( \xi \!\geq\! \omega') \\
\end{array}
\right.
\label{eq:Z-McM}
\\
N_{\rm c}(\xi) 
&=&
\!N_{\rm c}
,
N_{\rm a}(\xi)
\!=\!
\left\{
\begin{array}{ll}
\!-\!N_{\rm a}, & \  ( \xi \!\leq\!-\omega' ) \\
0,     & \  (|\xi| \!<\! \omega') \\
N_{\rm a}, & \  (\xi \!\geq\! \omega') \\
\end{array}
\right.
\label{eq:N-McM}
\\
\mathcal{K}_{\rm c}(\xi,\xi')
&=&
K_{\rm c}
, \nonumber \\
\mathcal{K}_{\rm a}(\xi,\xi')
\!&=&\!
\left\{
\begin{array}{ll}
\!-\! K_{\rm a}, & \  \bigl[( \xi \leq -\omega' \ \ {\rm and} \ \ \xi'\leq -\omega' ) \\
 & \ \ {\rm or} \ \ ( \xi \geq \omega' \ \ {\rm and} \ \  \xi' \geq \omega') \bigr] \\
K_{\rm a}, & \   \bigl[(\xi\leq -\omega' \ \ {\rm and} \ \  \xi' \geq \omega') \\
 & \ \ {\rm or} \ \ ( \xi \geq \omega' \ \ {\rm and} \ \ \xi'\leq -\omega' ) \bigr] \\
 0,     & \  \bigl(|\xi| < \omega' \ \ {\rm or}\ \ |\xi'| < \omega' \bigr) \\
\end{array}
\right.
.
\label{eq:K-McM}
\end{eqnarray}
Here, in addition to $\omega_{\rm D}$, we have also introduced $\omega'$ which specifies the energy scale where DOS and kernels substantially deviate from the values at $E_{\rm F}$. These forms are depicted in Fig.~\ref{fig:terms}.

Similarly to McMillan's procedure,\cite{McMillan} we assume the following rectangular form for the solution
\begin{eqnarray}
\Delta(\xi)
=
\left\{
\begin{array}{ll}
\Delta_{-}, & \ \  (-\omega' \geq \xi \geq -\omega_{\rm D}) \\
\Delta_{0}, & \ \  (|\xi| < \omega') \\
\Delta_{+}, & \ \  (\omega_{\rm D} \geq \xi \geq \omega') \\
\end{array}
\right.
.
\label{eq:gap-McM}
\end{eqnarray}
The condition that a nonzero solution for $\Delta$ exists gives $T_{\rm c}$. By retaining the lowest order term with respect to the values with subscript ``a", we obtain
\begin{eqnarray}
\!\!
T_{\rm c}
&\propto&
{\rm exp}\!
\left\{\!
\frac{1\!\!+\!\!Z_{\rm c}}{K_{\rm c}\!N_{\rm c}}
\!-\!
\frac{N_{\rm a}\!Z_{\rm a}{\rm ln}\!\left[\omega_{\rm D}/\omega'\right]}{N_{\rm c}(1+Z_{\rm c})}
\!+\!
\frac{Z_{\rm a}^{2}{\rm ln}\!\left[\omega_{\rm D}/\omega'\right]}{(1+Z_{\rm c})^{2}}\!
\right\}
.
\label{eq:Tc-McMillan}
\end{eqnarray}
Since $Z_{\rm a}$ is dependent on $N_{\rm a}$ through Eq.~(\ref{eq:Z-ave-new}), we can substitute an approximate form for $Z_{\rm a}$ in terms of $N_{\rm a}/N_{\rm c}$, and then we obtain
\begin{eqnarray}
T_{\rm c}
&\propto&
{\rm exp}
\left\{\!
-
\frac{1+\lambda}{\lambda}
-
\left(
\frac{N_{\rm a}}{N_{\rm c}}
\right)^{2}
\Lambda(r, \lambda)
{\rm ln}r
\right\}
.
\label{eq:Tc-McMillan2}
\end{eqnarray}
Here, $r\equiv\omega_{\rm D}/\omega'$, and we have used the following relation $K_{\rm c}N_{\rm c}=-\lambda$ and $Z_{\rm c}=\lambda$~(Ref.~\onlinecite{GrossI}). The first term in the bracket corresponds to the zeroth order, which appears in literatures.\cite{McMillan} The function $\Lambda(r,\lambda)$ is a positive function which monotonically increases by increasing $r$ or $\lambda$ and converges to a finite value $<1$ in the limit $r \rightarrow \infty$ and $\lambda \rightarrow \infty$. The detail of the derivation of Eqs.~(\ref{eq:Tc-McMillan}) and (\ref{eq:Tc-McMillan2}) is given in Appendix \ref{app:McM}.

From Eq.~(\ref{eq:Tc-McMillan2}), we see how the antisymmetric part affects $T_{\rm c}$: It reduces $T_{\rm c}$ regardless of the sign of $N_{\rm a}$ and its amount becomes substantial when the ratios $N_{\rm a}/N_{\rm c}$ or $r\equiv\omega_{\rm D}/\omega'$ are large. In addition, we find the antisymmetric part of the nondiagonal kernel ($K_{\rm a}$) does not contribute to $T_{\rm c}$ within the lowest order.

\subsection{Numerical calculation}\label{subsec:modelcalc}
The above analysis tells us that the asymmetry effect becomes pronounced for the cases where $N_{\rm a}/N_{\rm c}$ and $r$ are large. To quantify the effect more explicitly, we calculated $T_{\rm c}$ by numerically solving Eq.~(\ref{eq:gap-eq-ave}) using a model DOS. We included both the phonon and electron contribution to the kernels,  $\mathcal{K}
=
\mathcal{K}^{\rm ph}
+
\mathcal{K}^{\rm el}
$ and $\mathcal{Z}=\mathcal{Z}^{\rm ph}$. We employed the $n{\bf k}$-averaged form for $\mathcal{K}^{\rm ph}$ [Eq.~(23) in Ref.~\onlinecite{GrossII}], and we treated $\mathcal{K}^{\rm el}$ by an approximate form
\begin{eqnarray}
\mathcal{K}^{\rm el}(\xi,\xi')
=
\frac{\mu}{N(0)}
\end{eqnarray}
with a certain energy cutoff.  For $\mathcal{Z}^{\rm ph}$, we employed the previous form [Eq.~(\ref{eq:Z-ph-ave-prev})] and the new one [Eq.~(\ref{eq:Z-ave-new})].

\begin{figure}[b!]
\begin{center}
\includegraphics[width=8.5cm,clip]{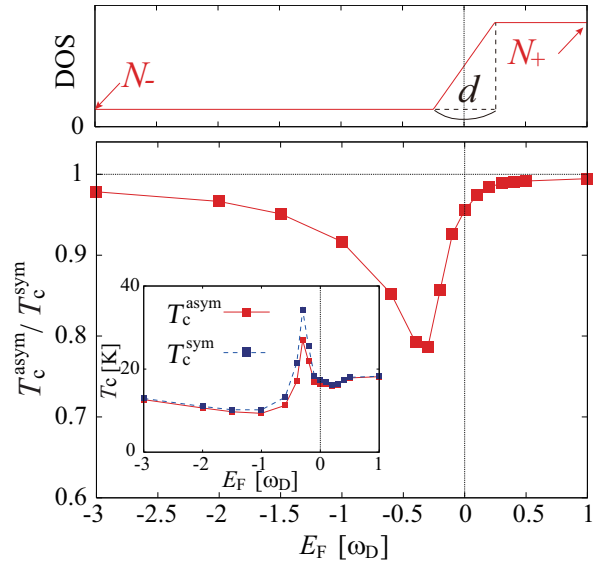}
\end{center}
\caption{(Color online) (Bottom) Ratio of $T_{\rm c}$ calculated using $\mathcal{Z}^{\rm ph, new}$ [Eq.~(\ref{eq:Z-ave-new})] ($T_{\rm c}^{\rm asym}$) to that calculated using $\mathcal{Z}^{\rm ph}$ [Eq.~(\ref{eq:Z-ph-ave-prev})] ($T_{\rm c}^{\rm sym}$) plotted as a function of $E_{\rm F}$. In the inset, absolute values of $T_{\rm c}^{\rm asym}$ and $T_{\rm c}^{\rm sym}$ are shown. (Top) DOS used for the calculations. We set $N(\xi)$$=$$N(E_{\rm F})$ for $|\xi|>5$eV.}
\label{fig:Tc-ratio}
\end{figure}

We introduced a step-like DOS as depicted in the upper panel of Fig.~\ref{fig:Tc-ratio} so that $N_{\rm a}/N_{\rm c}$ and $r$ can be large. This form is characterized by two parameters; the ratio of the high-energy and low-energy values $N_{+}/N_{-}$, and the energy range where DOS changes $d$. For the Eliashberg function, we used the following model function 
\begin{eqnarray}
\alpha^{2}F(\omega)
=
\frac{\lambda}{2\sigma \sqrt{\pi}}
\omega
{\rm exp}
\left[
-\biggl(\frac{\omega-\omega_{\rm D}}{\sigma}\biggr)^{2}
\right]
\label{eq:a2F-model}
\end{eqnarray}
with width $\sigma=\omega_{\rm D}/10$. We set $\omega_{\rm D}=400$[cm$^{-1}$], $d=\omega_{\rm D}/2$, and $N_{+}/N_{-}=6$. This setting is realistic in that one often see a similar dependence in doped semiconductors with quasi-2D Fermi surfaces.\cite{Felser-MNCl,Akashi-MNCl,Iwasa-MoS2}

With these settings, we calculated $T_{\rm c}$ for various settings of $E_{\rm F}$. The calculation was performed with a logarithmically uniform energy grid, where the minimum energy cutoff was set to $|\xi|=10^{-6}$ eV and the number of points per digit was 20. The energy cutoff for $\mathcal{K}^{\rm el}$ was fixed to 20eV. The parameters $\lambda$ and $\mu$ were fixed to $1.0$ and $0.5$ for all the setting of $E_{\rm F}$ so that the resulting $T_{\rm c}$ becomes $\gtrsim$ $10$K.

In the lower panel of Fig.~\ref{fig:Tc-ratio}, we plot the ratio of the calculated $T_{\rm c}$ using $\mathcal{Z}^{\rm ph, new}(\xi)$ [Eq.~(\ref{eq:Z-ave-new})] to that using $\mathcal{Z}^{\rm ph}(\xi)$ [Eq.~(\ref{eq:Z-ph-ave-prev})] as a function of $E_{\rm F}$, where the absolute values of $T_{\rm c}$ are given in the inset.\cite{note-Tc} The ratio is systematically less than unity; the effect of asymmetry lowers $T_{\rm c}$ for any $E_{\rm F}$, which is consistent with Eq.~(\ref{eq:Tc-McMillan2}). As the Fermi level approaches to the lower edge of the slope, the decrease of the ratio becomes significant and finally amounts to more than 20\%. When the Fermi level is located on the upper edge of the slope, the decrease is not appreciable. This $E_{\rm F}$ dependence can also be consistently understood with Eq.~(\ref{eq:Tc-McMillan2}); the characteristic energy scale ($\omega'$) decreases as the Fermi level approaches to the slope, and the ratio $N_{\rm a}/N_{\rm c}$ is relatively small when the Fermi level is on the upper edge. Thus, the change of $T_{\rm c}$ by the present improvement becomes crucial when $E_{\rm F}$ is close to a point where DOS increases rapidly. 

\begin{figure}[t!]
\begin{center}
\includegraphics[scale=0.60]{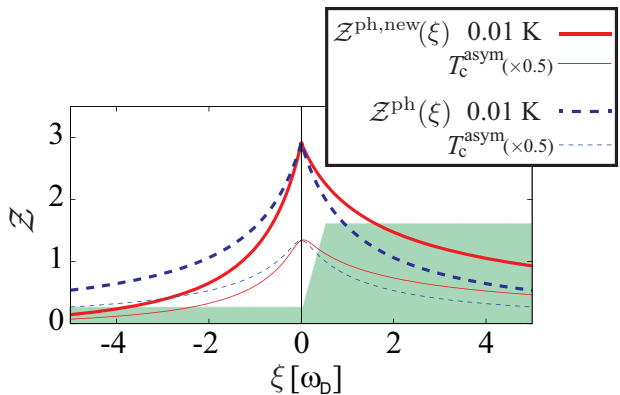}
\end{center}
\caption{(Color online) Calculated values of $\mathcal{Z}^{\rm ph, new}(\xi)$ (red solid line) and $\mathcal{Z}^{\rm ph}(\xi)$ (blue dotted line) for $E_{\rm F}$$=$$-0.3\omega_{\rm D}$. The green shaded area depicts DOS (arb. unit.) in the same energy scale. Note that the values for $T=T^{\rm asym}_{\rm c}$ (thin lines) are multiplied by 0.5.}
\label{fig:compare-Z}
\end{figure}

We show the calculated value of $\mathcal{Z}^{\rm ph}$ for $E_{\rm F}$$=$$-0.3\omega_{\rm D}$ in Fig.~\ref{fig:compare-Z} to examine the properties of the present form. The model DOS is also depicted in the same energy scale (green shaded area). For both $T$$=$$0.01$K (thick lines) and $T$$\simeq$$T^{\rm asym}_{\rm c}$ (thin lines), $\mathcal{Z}^{\rm ph, new}(\xi)$ becomes asymmetric in $\xi$, but they are similar to $\mathcal{Z}^{\rm ph}(\xi)$. In particular, the limiting value for $\xi \rightarrow 0$ agrees with $\lim_{\xi \rightarrow 0}\mathcal{Z}^{\rm ph}(\xi)$. These features well represent the properties (i)--(iii) described in the last paragraph of Sec.~\ref{subsec:derive-Z-new}. We observe a difference between $\mathcal{Z}^{\rm ph,new}(\xi)$ and $\mathcal{Z}^{\rm ph}(\xi)$ around $|\xi|\gtrsim \omega_{\rm D}$, which is responsible for the difference of $T^{\rm asym}_{\rm c}$ and 
 $T^{\rm sym}_{\rm c}$ in Fig.~\ref{fig:Tc-ratio}. Clearly, $\mathcal{Z}^{\rm ph, new}(\xi)$ becomes larger for the energy region with larger density of states, whereas $\mathcal{Z}^{\rm ph}(\xi)$ does not. The dependence of the former is reasonable because larger DOS should result in stronger mass renormalization due to stronger total electron-phonon coupling. Thus we were able to see that the particle-hole asymmetric electronic structure is properly treated with $\mathcal{Z}^{\rm ph,new}$ in a numerically stable manner.

Finally, let us comment on the application to actual superconductors. For typical superconductors, aluminum and niobium, we have 
carried out stable calculations of $T_{\rm c}$ (see appendix \ref{app:abinitio}), confirming again that the listed properties (i)--(iii) of $\mathcal{Z}^{\rm ph,new}$ are valid. The asymmetry effect on $T_{\rm c}$ was, however, estimated to be less than 0.1~\% for the both systems. This is mainly due to the small $N_{\rm a}/N_{\rm c}$ in these systems. The present weak asymmetry effect gives support to the validity of the particle-hole symmetrizing treatment\cite{Schrieffer, Mitrovic, Pickett-general, GrossI} for many cases of conventional superconductors.

\section{Summary and Conclusions}\label{sec:summary}
We constructed a new exchange-correlation kernel $\mathcal{Z}^{\rm ph,new}$ in SCDFT to treat the particle-hole asymmetry of electronic structure. The obtained $n{\bf k}$-averaged forms [Eq.~(\ref{eq:Z-ave-new})] and $n{\bf k}$-resolved form [Eq.~(\ref{eq:Z-new})] do not show any divergences or instabilities, and, for the systems with good symmetry, well agree with the previous symmetrized forms. By analytically deriving $T_{\rm c}$ formula, we found that the asymmetry systematically decreases $T_{\rm c}$. The amount of the decrease becomes substantial when the following two ratios are large: The ratio of the antisymmetic component of DOS to the constant component, and that of the Debye frequency to the characteristic energy scale of the DOS variation. We also calculated $T_{\rm c}$ with a model step-like DOS, showing that the amount of the reduction of $T_{\rm c}$ can be more than 20\%. With the present work, we successfully reinforced the theoretical foundation of SCDFT to extend its applicability to systems with significant particle-hole asymmetric electronic structure.

\begin{acknowledgments}
The authors thank Kazuma Nakamura, Shiro Sakai, Hideyuki Miyahara, and Jianting Ye for enlightening comments. This work was supported by Grants-in-Aid for Scientic Research from JSPS (No. 23340095), 
Funding Program for World-Leading Innovative R \& D on Science and Technology (FIRST program) on Quantum Science on Strong Correlation", 
JST-PRESTO and the Next Generation Super Computing Project and Nanoscience Program from MEXT, Japan.
\end{acknowledgments}
\begin{appendix}
\section{Evaluation of $\mathcal{Z}^{\rm ph}_{2}(\xi)$ for nonconstant DOS}\label{app:eval-Z2}
We here describe the detail of evaluation of $\mathcal{Z}^{\rm ph}_{2}(\xi)$ [Eq.~(\ref{eq:Z2-nonconst})] for the nonconstant DOS. Under the decomposition approximation of DOS introduced in Sec.~\ref{subsec:cancel}~[Fig.~\ref{fig:DOS-general}, Eq.~(\ref{eq:DOS-step})], $\mathcal{Z}^{\rm ph,2}(\xi)$ is given by 
\begin{eqnarray}
&&
\mathcal{Z}^{\rm ph,2}(\xi)
\nonumber \\
&&=
-
\biggl\{
\frac{1}{\xi}
-
\frac{\beta/2}{{\rm cosh}[(\beta/2)\xi]{\rm sinh}[(\beta/2)\xi]}
\biggr\}
\nonumber \\
&& \hspace{5pt} \times \!\!
\int \!\! d\omega
\alpha^{2}F(\omega)
\int^{L_{1}}_{-L_{2}}d\xi
\biggl\{
1 \!+\!
\sum_{i}[
N^{-}_{i}\theta^{-}_{i}(\xi)
\!+\!
N^{+}_{i}\theta^{+}_{i}(\xi)
]
\biggr\}
\nonumber \\
&& \hspace{10pt}
\times \!\!
\int^{L_{1}}_{-L_{2}}\!\! d\xi'
\biggl\{ \!
1 \!+\!
\sum_{j}[
N^{-}_{j}\theta^{-}_{j}(\xi')
\!+\!
N^{+}_{j}\theta^{+}_{j}(\xi')
]
\biggr\}
I'(\xi,\xi',\omega)
.\nonumber \\
\end{eqnarray}
With omitting the temperature-dependent terms of order $O[(T/\epsilon_{i})^{2}]$, $O(e^{-\beta \epsilon_{i}})$, $O(e^{-\beta L_{1}})$ and $O(e^{-\beta L_{2}})$, the integral for each term is carried out as 
\begin{eqnarray}
&&
\int^{L_{1}}_{-L_{2}} d\xi
\int^{L_{1}}_{-L_{2}} d\xi'
\theta^{\pm}_{i}(\xi)
\theta^{\pm}_{j}(\xi')
I'(\xi,\xi',\omega)
\nonumber \\
&&\simeq
\pm
{\rm ln}
\left|
\frac{(L_{1} \!+\! \epsilon_{i} \!+\! \omega)(L_{2} \!+\! \epsilon_{j} \!+\! \omega)}{(L_{1} \!+\! \epsilon_{j} \!+\! \omega)(L_{2} \!+\! \epsilon_{i} \!+\! \omega)}
\right|
,
\\
&&
\int^{L_{1}}_{-L_{2}} d\xi
\int^{L_{1}}_{-L_{2}} d\xi'
\theta^{\pm}_{i}(\xi)
\theta^{\mp}_{j}(\xi')
I'(\xi,\xi',\omega),
\nonumber \\
&&\simeq
\pm
{\rm ln}
\left|
\frac
{(L_{1} \!+\! \epsilon_{i}\!+\! \omega)(L_{1}\!+\! \epsilon_{j} \!+\! \omega)
(L_{2} \!+\! \epsilon_{i} \!+\! \omega)(L_{2} \!+\! \epsilon_{j} \!+\! \omega)}
{(L_{1} \!+\! L_{2} \!+\! \omega)^{2}(\epsilon_{i} \!+\! \epsilon_{j} \!+\! \omega)^{2}}
\right|
,
\nonumber \\
\\
&&
\int^{L_{1}}_{-L_{2}} d\xi
\int^{L_{1}}_{-L_{2}} d\xi'
\theta^{-}_{i}(\xi)
I'(\xi,\xi',\omega),
\nonumber \\
&&\simeq
{\rm ln}
\left|
\frac
{(L_{1} \!+\! L_{2} \!+\! \omega)^{2}(\epsilon_{i} \!+\! \omega)^{2}}
{(L_{1} \!+\! \omega)(L_{1} \!+\! \epsilon_{i} \!+\! \omega)
(L_{2} \!+\! \omega)(L_{2} \!+\! \epsilon_{i}\!+\! \omega)}
\right|
,
\\
&&
\int^{L_{1}}_{-L_{2}} d\xi
\int^{L_{1}}_{-L_{2}} d\xi'
\theta^{-}_{i}(\xi')
I'(\xi,\xi',\omega),
\nonumber \\
&&\simeq
{\rm ln}
\left|
\frac
{(L_{1} \!+\!\epsilon_{i} \!+\! \omega)
(L_{2} \!+\! \epsilon_{i} \!+\! \omega)}
{(L_{1} \!+\! L_{2} \!+\! \omega)^{2}}
\right|
,
\\
&&
\int^{L_{1}}_{-L_{2}} d\xi
\int^{L_{1}}_{-L_{2}} d\xi'
\theta^{+}_{i}(\xi)
I'(\xi,\xi',\omega),
\nonumber \\
&&\simeq
{\rm ln}
\left|
\frac
{(L_{1} \!+\! \epsilon_{i} \!+\! \omega)
(L_{2} \!+\! \omega)}
{(L_{2} \!+\! \epsilon_{i} \!+\! \omega)
(L_{1} \!+\! \omega)}
\right|
,
\\
&&
\int^{L_{1}}_{-L_{2}} d\xi
\int^{L_{1}}_{-L_{2}} d\xi'
\theta^{+}_{i}(\xi')
I'(\xi,\xi',\omega),
\nonumber \\
&&\simeq
{\rm ln}
\left|
\frac
{L_{2} \!+\! \epsilon_{i} \!+\! \omega}
{L_{1} \!+\! \epsilon_{i} \!+\! \omega}
\right|
.
\end{eqnarray}
Using these relations, $\mathcal{Z}^{\rm ph}_{2}(\xi)$ is transformed to the same form as Eq.~(\ref{eq:Z1-generalDOS}) with the opposite sign.

\section{Derivation of Eq.~(\ref{eq:Tc-McMillan}) and Eq.~(\ref{eq:Tc-McMillan2})}\label{app:McM}
Let us first plug the kernels and the gap function given in Eqs.~(\ref{eq:Z-McM})--(\ref{eq:gap-McM}) into the energy-averaged gap equation [Eq.~(\ref{eq:gap-eq-ave})].
The condition to have a nonzero solution is
\begin{widetext}
\begin{eqnarray}
{\rm det} M&=&0, \nonumber \\
M
&\equiv&
\left[
\begin{array}{ccc}
1 \!+\! Z_{\rm c} \!-\! Z_{\rm a}+\frac{1}{2}(N_{\rm c} \!-\! N_{\rm a})(K_{\rm c} \!-\! K_{\rm a})q &\ \  N_{\rm c}K_{\rm c}p &\ \ \frac{1}{2}(N_{\rm c} \!+\! N_{\rm a})(K_{\rm c} \!+\! K_{\rm a})q \\
\frac{1}{2}(N_{\rm c} \!-\! N_{\rm a})K_{\rm c}q &\ \ 1 \!+\! Z_{\rm c} + N_{\rm c}K_{\rm c}p &\ \ \frac{1}{2}(N_{\rm c} \!+\! N_{\rm a})K_{\rm c}q \\
\frac{1}{2}(N_{\rm c} \!-\! N_{\rm a})(K_{\rm c} \!+\! K_{\rm a})q &\ \ N_{\rm c}K_{\rm c}p &\ \ 1 \!+\! Z_{\rm c} \!+\! Z_{\rm a} + \frac{1}{2}(N_{\rm c} \!+\! N_{\rm a})(K_{\rm c} \!-\! K_{\rm a})q
\end{array}
\right]
,
\end{eqnarray}
\end{widetext}
with $p$ and $q$ defined by
\begin{eqnarray}
p\!=\!\!\!
\int_{0}^{\omega'} \!\!\!\! d\xi'
\frac{{\rm tanh}[(\beta_{\rm c}/2)\xi']}{\xi'}
,
q\!=\!\!\!
\int_{\omega'}^{\omega_{\rm D}} \!\!\!\! d\xi'
\frac{{\rm tanh}[(\beta_{\rm c}/2)\xi']}{\xi'}
.
\end{eqnarray}
Here $\beta_{\rm c}$ denotes the inverse of $T_{\rm c}$. In order to treat the lowest-order contribution, we keep the terms up to the second order with respect to the values with subscript ``a" as
\begin{eqnarray}
{\rm det}M
&=& 0 \nonumber \\
&\simeq&
(1 \!+\! Z_{\rm c})^{3}
+
[K_{\rm c}N_{\rm c}C \!-\! K_{\rm a}N_{\rm c}q](1 \!+\!Z_{\rm c})^{2}
\nonumber \\
&& \hspace{10pt}
+
[
-\! K_{\rm c}K_{\rm a}N_{\rm c}^{2}qC
\!-\!
K_{\rm c}N_{\rm a}Z_{\rm a}q
\!-\!
Z_{\rm a}^{2}
]
(1+Z_{\rm c})
\nonumber \\
&& \hspace{20pt}
-
K_{\rm c}N_{\rm c}Z_{\rm a}^{2}p
,
\end{eqnarray}
where $C\equiv p+q$ has been introduced. Retaining the lowest-order terms, we obtain
\begin{eqnarray}
C
&\simeq&
-
\frac{1+Z_{\rm c}}{K_{\rm c}N_{\rm c}}
+
\frac{N_{\rm a}Z_{\rm a}q}{N_{\rm c}(1+Z_{\rm c})}
-
\frac{Z_{\rm a}^{2}q}{(1+Z_{\rm c})^{2}}
.
\end{eqnarray}
Using
$
C={\rm ln}
\left[
2e^{\gamma}\omega_{\rm D} /(\pi T_{c})
\right]
$
with $\gamma$ being the Euler constant, we get Eq.~(\ref{eq:Tc-McMillan}).

Next we consider the relation between $Z_{\rm a}$ and $N_{\rm a}$. Let us start from our newly developed form [Eq.~(\ref{eq:Z-ave-new})]. The antisymmetric part of the integrand in $\xi'$ can be written as
\begin{eqnarray}
&&
\tilde{\mathcal{I}}(\xi,\xi',\omega)
-
\tilde{\mathcal{I}}(\xi,\xi',-\omega)
-
\tilde{\mathcal{I}}(\xi,-\xi',\omega)
+
\tilde{\mathcal{I}}(\xi,-\xi',-\omega)
\nonumber \\
&&
\hspace{10pt}
-\!
\tilde{\mathcal{J}}(\xi,\xi',\omega)
\!+\!
\tilde{\mathcal{J}}(\xi,\xi',-\omega)
\!+\!
\tilde{\mathcal{J}}(\xi,-\xi',\omega)
\!-\!
\tilde{\mathcal{J}}(\xi,-\xi',-\omega)
.\label{eq:antisym-part}
\nonumber \\
\end{eqnarray}
Using the facts $\tilde{\mathcal{I}}(-\xi,-\xi',-\omega)=\tilde{\mathcal{I}}(\xi,\xi',\omega)$ and $\tilde{\mathcal{J}}(-\xi,-\xi',-\omega)=\tilde{\mathcal{J}}(\xi,\xi',\omega)$, one can easily find the above part divided by ${\rm tanh}[(\beta/2) \xi]$ is antisymmetric in $\xi$. The symmetric part [Eq.~(\ref{eq:sym-part})] divided by ${\rm tanh}[(\beta/2) \xi]$ is, on the other hand, symmetric in $\xi$. Consequently, the antisymmetric part of $\mathcal{Z}^{\rm ph, new}(\xi)$ is yielded by only the antisymmetric part of  $N(\xi')$.

Substituting $N(\xi')=N_{\rm c}(\xi')+N_{\rm a}(\xi')$, for the antisymmetric contribution, we obtain
\begin{eqnarray}
&&\int d\xi'
\frac{N_{\rm a}(\xi')}{N_{0}}
[\mathcal{I}(\xi,\xi',\omega) \!-\! 2\mathcal{J}(\xi,\xi',\omega)]
\nonumber \\
&& \hspace{10pt}
=
2\frac{N_{\rm a}}{N_{\rm c}}
\int_{\omega'}^{\omega_{\rm D}}d\xi'
[(\textrm{antisym.})]
,
\end{eqnarray}
where (antisym.) represents the terms in Eq.~(\ref{eq:antisym-part}). This integration is easily performed since we can assume $\xi \gg T$, $\xi' \gg T$, and $\omega \gg T$, so that we get
\begin{eqnarray}
&&
2\frac{N_{\rm a}}{N_{\rm c}}
\int_{\omega'}^{\omega_{\rm D}}d\xi'
[(\textrm{antisym.})]
\nonumber \\
&& \simeq
2\frac{N_{\rm a}}{N_{\rm c}}
\int_{\omega'}^{\omega_{\rm D}}d\xi'
\frac{1}{(\xi+\xi'+\omega)^{2}}
\nonumber \\
&& =
2\frac{N_{\rm a}}{N_{\rm c}}
\left[
\frac{1}{\xi+\omega'+\omega}
-
\frac{1}{\xi+\omega_{\rm D}+\omega}
\right]
.
\end{eqnarray}
Finally, by using the Einstein spectrum $\alpha^{2}F(\omega)=\frac{\lambda}{2}\omega_{\rm D}\delta(\omega-\omega_{\rm D})$, we obtain
\begin{eqnarray}
&&
\frac{1}{2}
\left[
\mathcal{Z}^{\rm ph,new}(\xi) \!-\! \mathcal{Z}^{\rm ph,new}(-\xi)
\right]
\nonumber \\
&& \hspace{10pt}=
{\rm sgn}(\xi)
\lambda
\frac{N_{\rm a}}{N_{\rm c}}
\frac{\omega_{\rm D}(\omega_{\rm D}-\omega')}{(\xi \!+\! \omega' \!+\! \omega_{\rm D})(\xi \!+\! 2\omega_{\rm D})}
,
\label{eq:z-asym-def}
\end{eqnarray}
where the sign function comes from $\lim_{\beta \rightarrow \infty}$ $1/{\rm tanh}[(\beta/2)\xi]$. 

Subsequently, we consider to approximate this form as the rectangular form [Eq.~(\ref{eq:Z-McM})]. Using the value at $\xi=\omega_{\rm D}$ yields
\begin{eqnarray}
Z_{\rm a}
&=&
\lambda \frac{N_{\rm a}}{N_{\rm c}}
\frac{r-1}{3(2r+1)}
\nonumber \\
&\equiv&
\lambda \frac{N_{\rm a}}{N_{\rm c}}
h(r)
,
\end{eqnarray}
where $r$$=$$\omega_{\rm D}/\omega'$$>$$1$. By using this form in Eq.~(\ref{eq:Tc-McMillan}) and substituting $K_{\rm a}N_{\rm c}=-\lambda$, $Z_{\rm c}=\lambda$, we get to
\begin{eqnarray}
T_{\rm c}
&\propto&
{\rm exp}
\Biggl\{
-\frac{1\!+\!\lambda}{\lambda}
\nonumber \\
&&
-
\left(
\frac{N_{\rm a}}{N_{\rm c}}
\right)^{2} \!\!
\left[
\frac{\lambda h(r)}{1 \!+\! \lambda}
\!-\!
\left(
\frac{\lambda h(r)}{1\!+\! \lambda} 
\right)^{2}
\right]
{\rm ln}r
\Biggr\}
.
\label{eq:Tc-appendix}
\end{eqnarray}
This expression is equivalent to Eq.~(\ref{eq:Tc-McMillan2}), where the function $\Lambda(r, \lambda)$ is defined as the part in the square bracket in Eq.~(\ref{eq:Tc-appendix}). Its positiveness, monotonic dependences and the convergence in the limit $r \rightarrow \infty$ and $\lambda \rightarrow \infty$ referred in Sec.~\ref{subsec:analytic} are easily confirmed with this expression. We also note that $\Lambda(r, \lambda)$ has these properties for any values of $\xi$ ($\omega_{\rm D}$$>$$\xi$$>$$\omega'$) in Eq.~(\ref{eq:z-asym-def}) when we get an approximate form of $Z_{\rm a}$.

\section{{\it Ab initio} calculation}\label{app:abinitio}
\begin{figure}[b!]
\begin{center}
\includegraphics[scale=0.65]{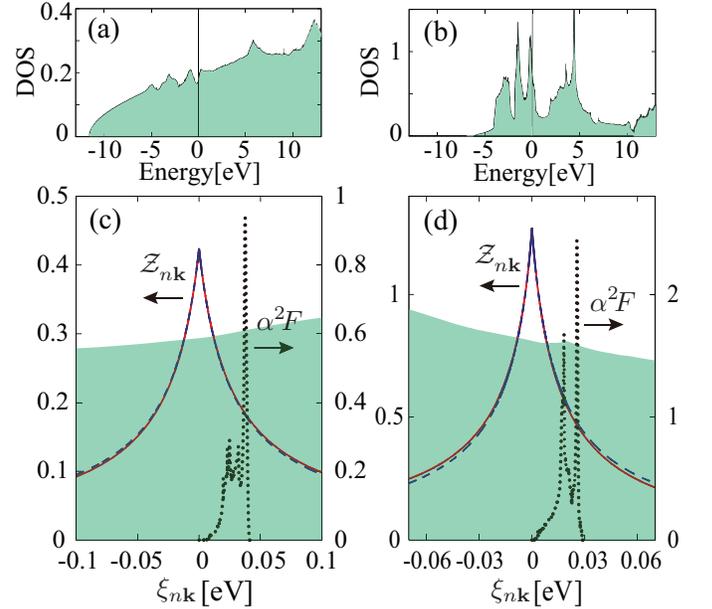}
\end{center}
\caption{(Color online) Calculated DOS for Al [(a)] and Nb [(b)]. $\mathcal{Z}$ kernels for Al [(c)] and Nb [(d)] calculated from Eq.~(\ref{eq:Z-ave-new}) (solid line) and Eq.~(\ref{eq:Z-ph-ave-prev}) (dashed line): The input Eliashberg function (dotted line) and DOS (green shaded area) are given in the same energy scale, with which $\lambda$ and $\omega_{\rm ln}$ were estimated for Al (Nb) as 0.417 (1.267) and 314~K (176~K).}
\label{fig:Al-Nb-Z}
\end{figure}
We applied our formalism to typical weak- and strong-coupling superconductors, Al and Nb. We solved the gap equation [Eq.~(\ref{eq:gap-eq})] with the energy-averaged approximation for the phonon kernels
\begin{eqnarray}
\mathcal{K}_{n{\bf k},n'{\bf k}'}
&=&
\mathcal{K}^{\rm ph}(\xi_{n{\bf k}},\xi_{n'{\bf k}'})
+
\mathcal{K}^{\rm el}_{n{\bf k},n'{\bf k}'}
,
\\
\mathcal{Z}_{n{\bf k}}
&=&
\mathcal{Z}(\xi_{n{\bf k}})
,
\end{eqnarray}
where $\mathcal{K}^{\rm ph}(\xi_{n{\bf k}},\xi_{n'{\bf k}'})$ and $\mathcal{K}^{\rm el}_{n{\bf k},n'{\bf k}'}$ are defined by Eq.~(23) in Ref.~\onlinecite{GrossII} and Eq.~(13) in Ref.~\onlinecite{Massidda}, respectively. For the diagonal part $\mathcal{Z}(\xi_{n{\bf k}})$, we employed $\mathcal{Z}^{\rm ph, new}(\xi_{n{\bf k}})$ [Eq.~(\ref{eq:Z-ave-new})] and $\mathcal{Z}^{\rm ph}(\xi_{n{\bf k}})$ [Eq.~(\ref{eq:Z-ph-ave-prev})]. We used an accurate random-sampling scheme given in Ref.~\onlinecite{Akashi-MNCl}, and the resulting sampling error was not more than a few percent. The detailed condition of the calculations is given in Ref.~\onlinecite{comment-detail}. 

In Fig.~\ref{fig:Al-Nb-Z}, the calculated DOS for Al [(a)] and Nb [(b)] are shown. The calculated values of $\mathcal{Z}_{n{\bf k}}$ with the two forms are also shown in panel (c) and (d) together with the input Eliashberg function $\alpha^{2}F$. For the both systems, the DOS (green shaded area) has small but nonzero antisymmetric component in the corresponding energy scale. However, the two forms of $\mathcal{Z}$ yield approximately the same value. For the calculated $T_{\rm c}$, regardless of the form of $\mathcal{Z}_{n{\bf k}}$, we obtained 0.7 and 8.7 K for Al and Nb, respectively. Although the gap values calculated with $\mathcal{Z}^{\rm ph, new}(\xi_{n{\bf k}})$ were approximately 0.1\% smaller than those calculated with $\mathcal{Z}^{\rm ph}(\xi_{n{\bf k}})$ for each set of the sampling points, this difference was within the sampling error. The present result indicates that the asymmetry effect is not significant for $T_{\rm c}$ in Al and Nb.

\end{appendix}


\begin{thebibliography}{999}
\bibitem{note-proof} We note that the condition $\epsilon_{i}$$\gg$$T$ must be retained when we take limit $N_{\epsilon}\rightarrow\infty$.
\bibitem{comment-subtract} We found that it is necessary to multiply $p(\xi'\pm\omega)$ also for the terms originating from $J_{2}$; otherwise, numerically $I_{1}$$-$$2J_{2}$ does not become zero at $\xi$$=$$0$ and consequently the kernel diverges for $\xi\rightarrow 0$.
\bibitem{note-Tc} The non-monotonic dependence of $T_{\rm c}^{\rm asym}$ and $T_{\rm c}^{\rm sym}$ on $E_{\rm F}$ is due to our approximation that $\lambda$ and $\mu$ are constant regardless of $E_{\rm F}$, and so we do not focus on that.
\end{thebibliography}
\end{document}